# Hund's coupling driven nature of magnetism in negative charge transfer material, SrCoO$_3$

Jyotsana Sharma[1,3],[*] Shivani Bhardwaj[1], and Sudhir K. Pandey[2][†]

[1]*School of Physical Sciences, Indian Institute of Technology Mandi, Kamand - 175075, India*
[2]*School of Mechanical and Materials Engineering,*
*Indian Institute of Technology Mandi, Kamand - 175075, India and*
[3]*Swami Vivekanand Govt. College Ghumarwin - 174021, India*
(Dated: December 14, 2025)

In this work, we investigate the microscopic origin of magnetism in SrCoO$_3$ by incorporating electronic correlations within the dynamical mean-field theory (DMFT) framework. We note a remarkable agreement of the calculated magnetic observables ( saturation magnetization ∼2.4 $\mu_B$; magnetic transition temperature, $T_c$∼350 K) with the experimental results. The system exhibits Hund's coupling-induced strong quasiparticle mass enhancements of upto $m^*/m$ ∼7 for Co 3$d$ states, with the largest renormalization occurring in the majority spin $t_{2g}$ orbitals, marking the onset of orbital-selectivity. Our results reveal a Stoner-*like* collapse of exchange splitting that drives the loss of long-range ferromagnetic order at $T_c$. The breakdown of Fermi-liquid behavior down to $T$∼100 K suggests a suppressed coherence scale. Local magnetic moment originates from a mixed-spin configuration formed through dynamical fluctuation between intermediate-spin and high-spin states. Large charge fluctuations ($\langle\Delta N^2\rangle$∼0.6) together with heavy quasiparticles establish the correlation effects regime, governed predominantly by Hund's physics in SrCoO$_3$.

## I. INTRODUCTION

The transition metal oxides belonging to the Ruddlesden-Popper (RP) series ($A_{n+1}B_nO_{3n+1}$) offer a variety of unconventional electronic and magnetic phenomena arising from strong correlations, reduced dimensionality, and interplay between lattice, charge, and spin degrees of freedom[1–6]. The end-member ($n = \infty$) pervoskites (ABO$_3$) are of particular significance as they represent the bulk three-dimensional limit of the RP series. The end-members of the homologous series majorly establish the foundation of the electronic structure, charge-transfer behavior, metal-ligand coupling etc. of the correlated-electron properties of the RP members with reduced dimensionality ($n$= 1,2,3..)[7–10].

The ABO$_3$ pervoskites have been extensively studied and found to exhibit phenomena of fundamental interest such as correlated metallicity, negative charge-transfer physics, spin-state transitions, structural transitions, ferroelectricity, colossal magneto-resistance, unconventional magnetism, superconductivity, large thermopower, spin-glass behaviour, metal-insulator transition etc.[11–18]. These end-members are largely comprised of systems that show paramagnetic metallic states (e.g., LaNiO$_3$, SrVO$_3$) or antiferromagnetic insulating states (e.g., LaFeO$_3$, LaMnO$_3$, SrMnO$_3$). However, among these, SrCoO$_3$ occupies a unique position as it remains cubic without any distortion and is one of the few transition-metal pervoskites -together with SrRuO$_3$- that stabilizes in a ferromagnetic ground state with remarkably high Curie temperature ($T_c$∼305 K)[19] for single-crystal sample, far exceeding its ruthenate counterpart ($T_c$∼160 K)[20].

The available experimental literature on the stoichiometric SrCoO$_3$ is limited primarily due to the difficulty associated with preparing samples with negligible oxygen deficiency[21, 22] . Nevertheless, the experimental measurements report that saturation magnetization (2-2.5 $\mu_B$) lies close to the intermediate-spin (IS) state ($S$ =3/2)[23], rather than the usually expected low-spin (LS) or high-spin (HS) ground state configurations. The earliest of studies proposed an IS ($S$ = 3/2) ground state (3$\mu_B$) as the natural explanation for the observed experimental magnetic moment of SrCoO$_3$, attributing it to the strong ligand hole coupling[24]. The majority of previous theoretical work on SrCoO$_3$ has been limited to static-mean field approaches[21, 23, 25], such as DFT or DFT+$U$, which fail to capture the dynamical fluctuations among LS, IS and HS multifolds and correlation effects, crucial for characterizing covalent Cobaltate systems. Moreover, a precise description of the effective spin state for the system resulting from the competition between the crystal field effects and Hund's coupling ($J$) would require a realistic account of $J$ (correspondingly, Hubbard $U$) since the crystal field limit is, to an extent, better captured by the static mean-field treatments.

The only study available with the dynamical many-body treatment (DFT+DMFT) for SrCoO$_3$ is by Kunes et al[24]., which claims the ground state effective spin of SrCoO$_3$ does not stem from a pure ionic configuration but rather from a superposition of different atomic spin states. They also propose a plausible double-exchange mechanism for magnetism, attributing it to the presence of a mixed-valence Co 3$d$ configuration. However, these results are obtained for a Hubbard $U$ value of ∼10 eV, which is unusually high for a metallic system. Their calculations show a remarkably high $T_c$ (∼1800 K), which is likely due to the density-density approximation of the Coulomb interaction form adopted in their work, as it suppresses the spin-flip and pair hopping terms essential for a realistic description of the spin fluctuations. The non-density-density terms become particularly more important for the accurate account of low-energy excitations and quasiparticle renormalization[26].

[*] jyotsana.sharma1290@gmail.com
[†] sudhir@iitmandi.ac.in

Furthermore, a recent experimental and DFT+DMFT investigation on highly strained $SrCoO_3$ thin films[27] revealed that ferromagnetism persists even across a strain-induced metal–insulator transition, indicating that the unexpectedly strong ferromagnetism observed in unstrained pure $SrCoO_3$, likewise originates from a correlation-driven mechanism rather than classical Zener double-exchange physics. These studies suggest the presence of an interesting magnetic phase in the $SrCoO_3$ driven by its spin state and strong covalence effects, as reported in the system, which includes negative charge transfer. Therefore, an advanced beyond-static meanfield treatment is necessary, considering the current advancements in the field, which include the accurate first-principles determination of Coulomb interaction parameters ($U$ and $J$) as well as the full Coulomb interaction matrix, to gain microscopic insights into the fundamental nature of its magnetic spin state and spectral properties. In this paper, we systematically examine the role of correlation effects in determining the magnetism and spectral properties of $SrCoO_3$, using first-principles-determined Hubbard $U$ and Hund's coupling $J$. We observe a Hund's-dominated description of magnetism present in the system, stabilizing in a mixed spin state configuration comprised primarily of IS and HS configurations. The system exhibits evidence of strong quasiparticle mass enhancements with a largely suppressed orbital-selective coherence scale.

## II. COMPUTATIONAL DETAILS

To understand electronic structure of $SrCoO_3$, we initially performed first-principle DFT calculations using APW+lo based WIEN2k code [28]. The PBEsol[29] exchange-correlation functional under GGA approximation has been used. The initial structural details were taken from XRD data for $SrCoO_3$[19]. The lattice parameter was optimised using volume optimisation tool in Wien2k. All calculations were performed over a mesh size of 12x12x12 keeping RMTKmax equal to 7. The RMT values for Sr, Co and O were taken as 2.30 au, 1.87 au and 1.61 au respectively. The convergence criteria was set for charge at $10^{-4} Coulomb/cell$. Both non magnetic (NM) and spin-polarised (SP) calculations were performed. The Gap2c[30] code is used for the constrained random phase approximation (cRPA) calculations to determine the partially screened Coulomb interaction parameters ($U$ and $J$) to be used for the DFT+DMFT calculations. Further, the value of $U$ has also been estimated via linear response method as implemented in the Abinit code[31][32]. The linear response method in Abinit computes $U$ by measuring the change in orbital occupation on applying small local potential shifts. The DFT+DMFT calculations have been performed using the eDMFT[33] code. Here impurity model of DMFT is solved for $U = 6.0$ eV and $J = 1.15$ eV using CTQMC (continous-time Quantum Monte Carlo) impurity solver. The results are obtained for various exact double-counting (DC) schemes[34] available in the eDMFTF code. The spin-rotational invariant (*full*) form of local Coulomb interaction is used for solving the DMFT impurity solver for realistic account of electronic interactions. A hybridization window ranging from -15 eV to 15 eV has been chosen for the DFT+DMFT calculations.

## III. RESULTS AND DISCUSSIONS

$SrCoO_3$ is known to possess simple cubic $Pm\bar{3}m$ structure[19]. The atoms Sr, Co and O are located at Wyckoff positions 1b (0.5,0.5,0.5), 1a (0,0,0) and 3d (0.5,0,0) respectively. The Co atom is surrounded by a regular octahedron of Oxygen ligands. This cubic octahedral geometry splits Co-3$d$ orbitals into triply degenerate $t_{2g}(d_{xy}, d_{yz}, d_{zx})$ and doubly degenerate $e_g(d_{3z^2-r^2}, d_{x^2-y^2})$ orbitals. To investigate the electronic structure in detail, we first performed lattice parameter optimization for our system. We calculated total energy of the system at different unit cell volumes. The total energy vs. volume per unit cell curve is shown in Fig. 1. This plot is then fitted with Birch-Murnaghan equation of state[35] to obtain volume corresponding to minimum total energy. The obtained volume from fitted curve is 364.41 $a_0^3$. This volume gave lattice constant value $a = 3.7798$ Å. The fitting of total energy vs. volume data also provides bulk modulus $B_0$ and its first pressure derivative $B_0'$. All these ground state material parameters calculated using PBEsol approximation are summarized in Table I alongwith results from previous theoretical and experimental works. Our calculated values agree fairly with the theoretical works of Ali & Ahmad[36] performed using Wu-Cohen GGA. On comparing the experimental value of lattice constant, 3.8289 Å[19] taken at room temperature, with our calculated value of 3.7798 Å which is at 0 K; we get a thermal expansion coefficient of $\sim 4.2 \times 10^{-5}$ K$^{-1}$. This value of thermal expansion coefficient agrees with those from previous work[37]. Hence, we can say our calculated value of lattice constant is consistent with the experimental results. Thus, the comparison of our present work with the previous works, ascertains that PBEsol exchange correlational functional works reliably for this compound. We have used the volume-optimized calculated value of lattice parameter in this study throughout.

From DFT calculations, total magnetic moment came out to be 2.50 $\mu_B$/f.u.. The magnetic moment contribution from within muffin tin of Co is 1.83 $\mu_B$,

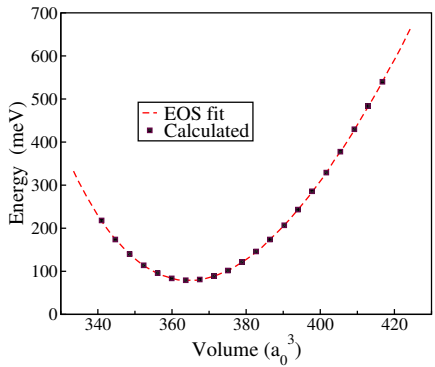

FIG. 1: Energy vs. Volume per unit cell

TABLE I: Calculated material parameters lattice constant $a$, bulk modulus $B_0$ and its first pressure derivative $B_0'$ for SrCoO$_3$ compared to previous theoretical and experimental results

| Parameters | Present Work | PreVious Theoretical | Experimental |
|---|---|---|---|
| $a$ (Å) | 3.7798 | 3.78[36] | 3.8289[19] |
| | | | 3.8246[38] |
| $B_0$ (GPa) | 173.57 | 176.14[36], 144[11] | - |
| $B_0'$ (GPa) | 4.95 | 5.32[36], 4[11] | - |

Ref.[36] used Wu-Cohen GGA [39] as implemented in WIEN2K
Ref.[11] used GGA as implemented in VASP[40]

TABLE II: Total magnetic moment of SrCoO$_3$ compared to previous theoretical and experimental results

| Parameters | Present Work | Previous Theoretical | Experimental |
|---|---|---|---|
| $\mu_{total}$ ($\mu_B$/f.u.) | 2.50 | 2.45[36], 2.28[21] | 2.5[19] |
| $\mu_{Co}$ ($\mu_B$/atom) | 1.83 | 1.79[36], 1.69[25] | 1.78[42] |
| $\mu_O$ ($\mu_B$/atom) | 0.18 | 0.17[36], 0.19[25] | - |
| $\mu_{interstitial}$ ($\mu_B$) | 0.13 | 0.15[36] | - |

Ref.[36] used Wu-Cohen GGA [39]
Ref.[21] used GGA+U(with $U_{eff} = 5eV$)
Ref.[25] used von Barth−Hedin exchange correlation

O is 0.18 $\mu_B$ and interstitial is 0.13 $\mu_B$. The finite magnetic moment on Oxygen atoms indicates the possibility of negative charge transfer[23, 41] from O-2$p$ to Co-3$d$ resulting in unpaired electron and hence, the covalent contributions to the bonding in SrCoO$_3$. The experimental value of low temperature moment for single crystal stoichiometric SrCoO$_3$ at 2 K is 2.5 $\mu_B$/f.u.[19] when 7 T magnetic field is applied. The calculated magnetic moment value comes out to be in good agreement with the experimental[19] and previous theoretical[21, 25] results (tabulated in Table II).

Since SrCoO$_3$ is a strongly correlated system, we employed DFT + DMFT approach to incorporate correlation effects into our study of the magnetization of SrCoO$_3$ as a function of temperature. The predictive power of DMFT depends majorly on Coulomb interaction parameters $U$ and $J$ as these parameters account for strength of correlations among electrons. Different DFT+$U$ studies have employed various $U$ values for SrCoO$_3$ such as 2.5[43] eV, 2.75[44] eV, 7.0[45] eV. We found only one DFT+DMFT study by Kunes et. al. [24], where $U = 10.83$ eV and $J = 0.76$ eV has been used but this value is too high for a metallic system. In this work we performed constrained random phase approximation (cRPA) calculations to determine the Coulomb interaction parameters $U$. Since cRPA has different tweaking parameters and the values of Coulomb interaction parameters are known to be sensitive to various factors such as the choice of correlated energy window, number of correlated bands, choice of projectors used for constructing Wannier functions etc.[46, 47], therefore, we have studied the variation in the values obtained depending on the number of bands constituting the correlated subspace to determine realistic $U$ for the DFT+DMFT calculations. The resulting values are tabulated in Table III, where $U_{full}$ ($U_{diag}$) denote the averages of the all (diagonal) matrix elements of the effective Coulomb interaction matrix. Firstly, all 14 bands in the energy window ranging from -7.5 eV to 4 eV, are selected for constructing maximally-localized Wannier functions, because, strong hybridization of Co-$d$ and O-$p$ states leads to largely entangled $d$-character bands as shown in Fig.2. Then we performed calculations with different number of bands masked to constrain the intra-3$d$ transitions. This results in large value of $U_{full} \sim 12$ eV and a similar order of $U_{diag}$. However, further excluding the 1st and 2nd bands from the correlated subspace leads to largely reduced $U$ ( $U_{full} \sim 7$ eV and $U_{diag} \sim 8$ eV) suggesting large screening effects from the 1st and 2nd bands. With further exclusion of the bands and considering the range of constrained bands as 4-14, 5-14 and 6-14, results in reduction of $U$ ($U_{full}$) values to ~6.16 eV, 5.41 eV and 4.86 eV, respectively. Note that the range selected with bands 4-14, is characterized by dominant Co 3$d$ character relative to other bands and yield a value of $U$ around 6 eV. Furthermore, we find that this value comes in close agreement with the linear response calculated $U$ ($U$=5.95 eV) (refer Table III). We therefore used $U$= 6.0 eV in the DFT+DMFT calculations with $J$ (correspondingly deduced from Yukawa screening method) = 1.15 eV for SrCoO$_3$.

We next examined the robustness of our results against the choice of different DC schemes ( i.e. *exactd*, *exact*,

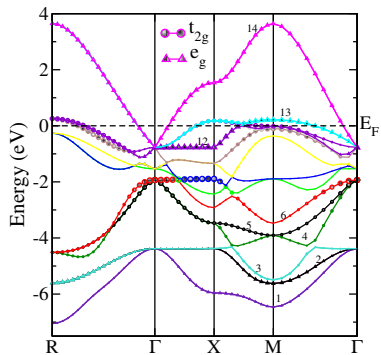

FIG. 2: Band character plot of Co-3d orbitals in NM phase obtained at DFT level.

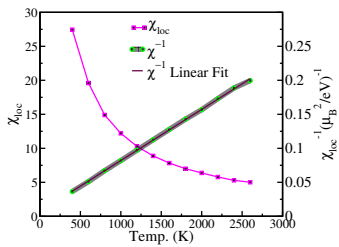

FIG. 3: Temperature dependent local spin susceptibility ($\chi_{loc}$) obtained in the paramagnetic phase above $Tc$ along with the inverse $\chi_{loc}$ and its Curie-Weiss fit.

TABLE III: ab-initio calculation of Coulomb interaction parameters. $U$ in the units of $eV$

| ab-initio method | index of bands masked | $U_{full}$ | $U_{diag}$ |
|---|---|---|---|
| cRPA | 1-14 | 12.20 | 12.54 |
|  | 3-14 | 7.02 | 8.2 |
|  | 4-14 | 6.16 | 7.41 |
|  | 5-14 | 5.41 | 6.63 |
|  | 6-14 | 4.86 | 6.06 |
| Linear Response | - | 5.95 | - |

TABLE IV: Values of saturation magnetization obtained using different dc schemes for $U_{full} = 6$ eV

| dc scheme | $\mu_B$/f.u. |
|---|---|
| exactd | 1.68 |
| exact | 1.88 |
| exacty | 2.44 |

*exacty*) by evaluating the variation in saturation magnetization value as given in Table. IV). Among the given DC schemes, *exacty* gives the value of saturation magnetization ~2.44$\mu_B$, exceptionally close to the experimental value of 2.50 $\mu_B$[19]. Moreover, the *exacty* DC scheme provides the magnetic transition temperature ($T_c$) of ~350 K, which lies in remarkable match with the experimental value of 305 K. This is consistent with the previous results of the metallic systems for which *exacty* is found to be more suitable DC scheme[5, 34]. Thus, *exacty* DC scheme is adopted throughout the study.

Fig. 3 depicts the local spin susceptibility ($\chi_{loc}$) and its inverse ($\chi_{loc}^{-1}$), in the paramagnetic phase above $T_c$. The $\chi_{loc}^{-1}$ shows an appreciably linear behavior over a large range of temperature (500-2500 K), indicating the presence of unscreened local moments in the paramagnetic phase of SrCoO$_3$. We find the local magnetic moment from the linear slope of $\chi_{loc}^{-1}$ by fitting as:

$$\chi_{loc}^{-1} = 3(g\mu_B)^2(T+T_w)/\mu_{loc}^2$$

where $g$= 2, and $T_w$ is the Kondo temperature.

The fit yields a local moment value of ($\mu_{loc}$ ~3.21 $\mu_B$, representing an effective spin state of $S = 1.2$ (2.4 $\mu_B$). This estimated value comes in quite good agreement with the observed saturation magnetization of SrCoO$_3$ in the experimental reports (~2.5 $\mu_B$). The overall remarkably good agreement of the calculated $T_c$ and the low temperature magnetization (2.44 $\mu_B$) from the single-site DMFT, reflects a Hund's dominated description of magnetism in this system.

Next, the effect of electronic correlations on the spectral properties of SrCoO$_3$ is presented. Fig. 4 (i) shows spin-orbital resolved DFT+DMFT PDOS computed at 110 K alongwith DFT PDOS. The inclusion of correlation effects introduces notable changes with respect to the DFT. Firstly, for all Co-3d states, new incoherent features appear at higher binding energies (~8-10 eV below $E_F$) where DFT PDOS is negligible and the low-energy quasiparticle states become narrower. We can observe that peaky structures form in the DFT+DMFT PDOS for both $t_{2g}$ and $e_g$ orbitals, marking significant renormalization of the DFT PDOS due to correlation induced smearing of states. The system exhibits an orbital-dependent polarization at $E_F$ with $t_{2g}$ states dominating the minority and $e_g$ orbitals dominating the majority spin-channels. The $t_{2g}$-up states show a gap formation (Mott-like localization) in the DFT+DMFT PDOS (refer inset of part (a) of Fig. 4 (i)), as incoherent weight starts to appear above the $E_f$. On the contrary the DFT PDOS of $t_{2g}$-up states lie almost fully occupied below the $E_f$ (clearly visible from the absence of DFT PDOS in very close proximity of the $E_F$:inset of part (a) Fig 4 (i)). The $t_{2g}$-dn sector shows quite small renormalization of DFT PDOS, which is evident from a similar spectral weight distribution as that of its DFT PDOS around the $E_f$, also relatively robust bandwidth against correlations and thermal effects (refer figure 1. of supplemental material). Here the results point to majorly localized nature of the $t_{2g}$ sector. The $e_g$ orbitals display formation of major quasiparticle spectral weight around the $E_F$ in both their majority and minority spin channels, resulting from their itinerant character. The low energy normalization of PDOS obtained after treatment with dynamical correlations suggest evolution of DOS into a spin-orbital selective Mott-*like* localization at low temperature.

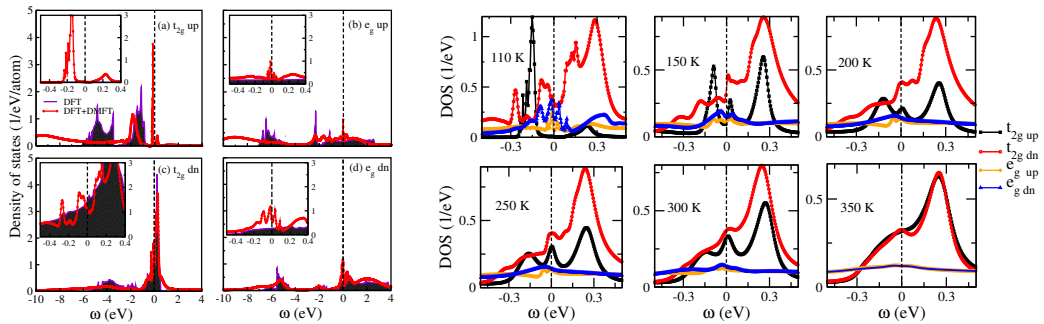

FIG. 4: (Right)(i) Spin and orbital resolved DOS plot obtained from DFT (shaded) and DFT+DMFT at 110K (unshaded) for $t_{2g}$ and $e_g$ orbitals. Dashed line at $\omega = 0$ represents $E_f$. (Left)(ii) Temperature-dependent evolution of the spin–orbital resolved DOS below $T_c$ in the vicinity of $E_f$ at DFT+DMFT level.

Further, to study the evolution of quasiparticle spectral weight distribution across the FM phase, its temperature dependent evolution of spectral density of Co $3d$ states is given in Fig. 4 (ii). The presence of quasiparticle peaks is noted at low temperatures $T{\sim}110$ K for the $e_g$ sector, which eventually lose coherence upon increasing the temperature to 200 K and above, suggesting a coherence scale of around 100 K. While the majority $t_{2g}$ sector is localized and minority sector shows largely broadened quasiparticle peak at $E_f$, suggesting even lower temperature coherence scale than 110 K. As the temperature is increased both the $t_{2g}$ and $e_g$ sectors undergo significant thermal broadening of the quasiparticle states, where the $e_g$ sector loses coherence more rapidly among the two, which is visible from the significant broadened weight of $e_g$ states at $T=150$ K. Correspondingly, the Hund's exchange splitting in the $e_g$ manifold collapses ${\sim}250$ K, significantly earlier than in the $t_{2g}$ channel, where the exchange splitting persists upto $T_c$. Interestingly, at 150 K the $t_{2g}$-dn orbital shows a diverging behavior at the $E_f$, due to a possible presence of Van-Hove singularity at finite temperature, signature of which is also present as slight increase in its scattering rate value (which will be discussed later in the text; refer Fig. 5. Also the peak of $t_{2g}$-up DOS at ${\sim}-0.15$ eV splits into two peaks, leading to disappearance of gap at the $E_f$ which appears in the form of a pseudogap like feature, as the temperature is increased to 150 K. Overall, the exchange splitting, dominated in magnitude by the $t_{2g}$ states, decreases approximately in proportion to the net magnetization. The principal collapse of exchange splitting sets in around ${\sim}300$ K, close to the experimental Curie temperature of $SrCoO_3$. This seems consistent with the Stoner-like description of the itinerant ferromagnetic transition. Notably, this behavior contrasts with that of $SrRuO_3$ where finite exchange splitting survives even at the magnetic transition temperature[20], suggesting more degree of itinerant nature of magnetism in $SrCoO_3$ than its ruthenate counterpart.

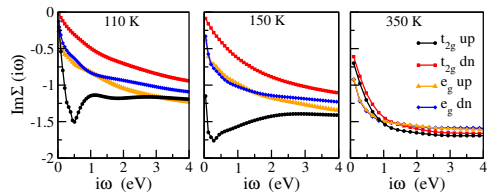

FIG. 5: Variation of spin-orbital resolved imaginary part of self-energy ($Im\Sigma(i\omega)$) in Matsubara frequency below $T_c$.

Next, the quasiparticle broadening of the Co $3d$ states is quantified by the imaginary part of self-energy obtained in the Matsubara frequency regime. In Fig.5, the temperature evolution of imaginary parts of self-energy ($Im\Sigma(i\omega)$) of the $t_{2g}$ and $e_g$ orbitals are shown on the imaginary frequency axis. It is clearly seen that at low energies, the behavior of $Im\Sigma(i\omega)$ is qualitatively different for $t_{2g}$ and $e_g$ orbitals and there is further level of spin-polarized distinction within the $t_{2g}$ manifold. This also marks the onset of orbital selective Fermi-liquid coherence scale in the system, a characteristic signature of Hunds induced correlations. At low temperatures ${\sim}110$ K, while the $Im\Sigma(i\omega)$ of $t_{2g}$-dn state shows linearly vanishing behavior at low-frequency region consistent with Fermi liquid behavior, the $t_{2g}$-up states exhibit a diver-gent behavior (non-linear) near the low imaginary frequency region ($i\omega$= 0-1 eV) indicating strong deviation from the Fermi-liquid behavior and signature of Mott-*like* localization of the $t_{2g}$-*up* state, as discussed in the earlier section. The $e_g$ orbital shows nearly Fermi-liquid *like* behavior in both its majority and minority spin-channels with quite small values of scattering rate ($\sim$0.15 eV) for each channel at 110 K. As the temperature is increased to 150 K, the behavior remains same except that the $Im\Sigma(i\omega)$ curve of the $t_{2g}$-*dn* state gains finite intercept towards zero Matsubara frequency showing finite quasiparticle lifetime and slightly increased scattering rate. However, the $t_{2g}$-*up* state shows an enlarged scattering rate at 150 K ($\sim$1.2 eV). On the other hand, the scattering rate of $e_g$ orbital for both spin-*up* and spin-*dn* channels remain relatively unaffected as the temperature is increased to 150 K. However, around $T_c$ at $T$= 300 K, the curves of $Im\Sigma(i\omega)$ of both the spin channels of $t_{2g}$ orbitals overlap at $\sim$0.65 eV, which results from reduced scattering rate value for the $t_{2g}$-*up* channel resulting from appearance of finite spectral weight at $E_f$ after disappearance of gap and increased scattering rate for the $t_{2g}$-dn channel, due to increased temperature dependent incoherence in $t_{2g}$-dn channel. Similarly, overlap in values is noted for the $e_g$-*up* and $e_g$-*dn* states at slightly large intercept ($\sim$1 eV). This is likely due to collapse of spin-polarization.

Correspondingly, the mass enhancement factor $Z^{-1}$, which is the ratio of effective mass ($m^*$) of interacting electron to the mass of non-interacting electron ($m_{DFT}$), is calculated as :

$$Z^{-1} = 1 - [\delta Im\Sigma(i\omega)/\delta\omega]_{0^+} \quad (1)$$

Table V. shows spin-orbital resolved mass enhancements for Co 3$d$ states at the lowest calculated temperature i.e. 110 K against the variation in the strength of $J$ (0, 0.6, 1.15 eV). At $J=$ 0 eV, the $t_{2g}$ and $e_g$ sector show nearly coherent character with $Z\sim$0.7-0.8, with slight mass enhancement for each sector. While increasing the $J$ value to 0.6 eV, a modest differentiation between the orbitals begins, especially the $t_{2g}$-*up* ($Z^{-1}$=2.44) and $eg$-*up* ($Z^{-1}$=2.00) sector gains relatively moderate mass enhancement relative to their spin-dn counterparts. However, at $J$=1.15 eV, which is the value of Hunds coupling benchmarked for the system, we observe drastic rise in the mass enhancement factor ($m^*/m_{DFT}$=$\sim$ 6-7) characterized by strong orbital differentiation. (Note that from the above discussion of the scattering rate behavior at $J$= 1.15 eV, the $t_{2g}$-*up* states show non-Fermi liquid behavior even at 110 K and hence quasiparticle picture breaks down. Therefore, for the $t_{2g}$-*up* states the first few Matsubara frequencies are considered for calculating the $Z^{-1}$, i.e. first six Matsubara frequency points, where the $Im\Sigma(i\omega)$ can be linearly approximated). Interestingly, the largest mass renormalization $Z^{-1}$ ($m^*/m_{DFT}$) = 7.13 is obtained for the $t_{2g} - up$ states. Whereas, the spin-*dn* counterpart of the $t_{2g}$ states, undergoes a relatively small mass renormalization of $\sim$1.88, which is expected since this channel experiences relatively less effective correlation induced renormalization due to large bandwidth and subsequent Fermi-liquid like behavior as mentioned in earlier. The $e_g$ sector also shows large mass enhancement values in both its spin-channels ($e_g$-*up*$\sim$6.24; $e_g$-*dn*$\sim$5.46), which is surprisingly nearly same as that of the $t_{2g}$-*up* states, inspite of largely Fermi-liquid behavior at 110 K.

Unlike $t_{2g}$-*up* states, the $t_{2g}$-*dn* sector majorly stays robust against the variation in $J$, which reflects spin-polarized nature of correlation effects defining the $t_{2g}$ sector. The strong quasiparticle mass renormalization for the $t_{2g}$-*up* (> 6) and $e_g$ (> 5) states at $J$= 1.15 eV result from Hunds-driven correlations in the system. These values are significantly larger than the mass enhancement observed for its Ruthenate counterpart i.e. SrRuO$_3$ ($m^*/m_{DFT} \sim$ 4), where it is attributed to strong coupling to the low-energy bosonic modes arguing weaker electronic correlation effects[20]. However, here in the case of SrCoO$_3$, the large mass enhancement results from electronic correlations. We also quantify the charge fluctuations ($\langle \Delta N^2 \rangle$) at $J$=1.15 eV and find the value to be $\sim$0.62 at 110 K and remains almost the same throughout the FM phase showing temperature independent variation. The substantial value of $\langle \Delta N^2 \rangle$ suggests a predominantly competing degree of itinerancy of Co 3$d$ electrons as also observed for recently reported value $\sim$0.6 for n=1 member of RP series i.e. Sr$_2$CoO$_4$[5]. The large orbital differentiated intermediate mass enhancement values accompanied by significant charge fluctuations mark onset of Hunds metal characteristics in multiorbital systems[48–50].

We further present the finite temperature atomic-multiplet probabilities of the Co 3$d$ states (Fig.6) and note that the occupation probabilities remain nearly temperature independent. The Co 3$d$ orbitals show a mixed-valence state with the largest probability corresponding to the 3$d^6$ multiplet, suggesting the ground state with large probability of the $d^6$ occupancy. The $d^5$, $d^7$ multiplets also show significantly finite probabilities, with $d^7$ being $\sim$60% of the $d^6$ probability leading to a mixed valence state indicating large covalency effects resulting from strong ligand hybridization. Interestingly, the presence of mixed-valence state of Co 3$d$ orbital has been previously also noted by Kunes et. al.[24], proposing a plausible double-exchange (DE) mechanism of ferromagnetism in the system. However, our results using *full* rotationally invariant form of Coulomb interaction do not support DE-driven ferromagnetism in SrCoO$_3$, since we obtain a qualitatively different picture of low-energy coherence in Co-3$d$ manifold. For DE to be relevant, the orbitals which mediate the hopping (typically the $e_g$ states) should maintain sufficient quasiparticle coherence for ensuring spin alignment between the neighboring Co (impurity) sites. On the contrary, our results show strong quasiparticle mass renormalization leading to largely suppressed $Z$ (<0.2) factor for both the spin-channels of $e_g$ orbitals. These $Z$ values indicate the presence of incoherent and heavy carriers. Therefore, the long-range coherent hopping required for a robust DE mechanism is expected to be suppressed. This assessment is further evident from the spin-orbital resolved scattering rates (from the $Im\Sigma(i\omega)$), which remain large at the lowest calculated temperature $T$=110 K for both the spin channels of $e_g$ orbitals. Therefore, the mixed-valence in SrCoO$_3$ signals strong covalency effects rather than a functional DE mechanism.

We systematically examine the electronic occupancies of the $t_{2g}$ and $e_g$ states to establish the role of correlation effects in deciding the spin-state configuration of the Co 3$d$ states in SrCoO$_3$. Table VI shows the orbital and spin resolved occupancies of the Co 3$d$ states obtained at the DFT level and also the occupancies obtained on employing the dynamical correlation effects using DFT+DMFT (at 110 K). The degree of correlations can be found based on the partial charges of Co 3$d$ electrons within DFT+DMFT in comparison to DFT. The DFT calculated occupancy of the Co 3$d$ orbital adds upto ∼6.41, whereas, in the DFT+DMFT calculations it is found to be ∼6.27 (at $J$=1.15 eV). This modest reduction of the partial charge by ∼0.14$e$ hints to a moderately correlated regime with not so strongly suppressed charge-fluctuations, consistent with the above noted conclusion from $\langle \Delta N^2 \rangle$ value. This stands in contrast to the previously reported large partial charge renormalization upto ∼1$e$, which should be interpreted as strong suppression of charge fluctuations. We note that this primarily arises from the use of large Hubbard $U$ (10 eV) together with density-density type of Coulomb interaction used in their work. This combination of -large $U$ and absence of non-density-density terms- is expected to strongly enhance charge localization by suppressing dynamical charge fluctuations and drive large spectral weight transfer from the quasiparticle states[26]. Consequently, neglecting the non-density-density terms tends to strongly localize the low-energy states and overestimate the loss of partial 3$d$ occupancy. Unlike their study, the smaller renormalization in our calculations using the full type with realistic $U$ indicates a Hunds metal like regime rather than strongly localized charge-frozen state.

Correspondingly, the effect of increase in the strength of Hund's coupling i.e $J$=0.0 eV to 1.15 eV, is given on respective occupations in Table VI.

It can be inferred that the DFT calculations yield an approximate IS state ($t_{2g}^{4.62}e_g^{1.79}$) and even $J$=0 eV yields a pure IS state within the DFT+DMFT framework. When the Hund's coupling strength is increased to 1.15 eV the 3$d$ configuration shows a finite contribution of the HS state as can be noted from the occupation, which is expected as Hund's coupling tends to favor maximum spin state ($t_{2g}^{4.06}e_g^{2.21}$) yielding a tendency towards HS configuration. Furthermore, with a strong Hund's coupling magnitude inherent to the system i.e. $J$= 1.15 eV, the analysis of its spin-state probability distribution suggests presence of mixed valence and mixed spin-state configuration at 110 K- as histogram peaks near $d^6$ multiplet with substantial $d^5$ and $d^7$ weights, while the spin-state distribution shows broad dominant contributions from S=1/2, 1 & 2 ( refer Fig 6. This suggests that magnetic moment does not arise from a single static spin-configuration but from a dynamical superposition of different atomic multiplets with an overall HS state favored by strong Hund's coupling ($J$= 1.15 eV) in the system. The mixed valence state with finite contributions from the several spin state multiplets suggests enhanced metal-ligand hybridization and spin-charge coupling in stabilizing the mixed-spin ground state. This is consistent with a recent cluster-mulitplet study by Zheng et al, while studying the effect of epitaxial strain on SrCoO$_3$ thin films[51]. We note that at low temperatures the system exhibits characteristics suggestive of the Hund's metallicity by displaying an intermediate correlations regime with both the itinerant and localized limits. This is particularly evident from the large $\langle \Delta N^2 \rangle$ along with strong mass enhancements pointing to presence of heavy quasiparticles down to the lowest calculated temperature, 110 K.

TABLE V: Spin-orbital resolved mass-enhancement factor ($m^*/m_{DFT}$) along with quasiparticle spectral weight factor ($Z$) given in bracket, at $T$= 110 K for $J$ = 0, 0.6 & 1.15 eV

| $J$ (eV) | $t_{2g}-up$ | $e_g-up$ | $t_{2g}-dn$ | $e_g-dn$ |
|---|---|---|---|---|
| 0 | 1.45 (0.68) | 1.27 (0.78) | 1.41 (0.70) | 1.27 (0.78) |
| 0.6 | 2.44 (0.40) | 2.00 (0.50) | 1.83 (0.54) | 1.89 (0.53) |
| 1.15 | 7.13 (0.14) | 6.24 (0.16) | 1.88 (0.53) | 5.46 (0.18) |

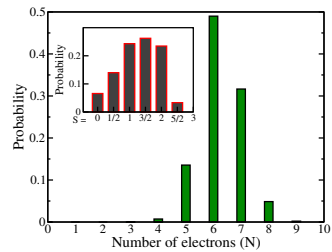

FIG. 6: Occupation probabilities of 3$d$- states at $T$= 110 K for $J$=1.15 eV. Inset displays probability distribution for different spin states at 110 K

TABLE VI: Orbital resolved occupancies of Co $3d$ states obtained using DFT and DFT+DMFT (at $T = 110$ K) methods

|  |  | $t_{2g}$ | $e_g$ |
|---|---|---|---|
| DFT |  | 4.62 | 1.79 |
| DFT+DMFT | $J = 0.00$ | 5.00 | 1.10 |
|  | $J = 0.60$ | 4.67 | 1.89 |
|  | $J = 1.15$ | 4.06 | 2.21 |

## IV. CONCLUSION

In conclusion, a realistic account of the electronic correlations within the dynamical mean-field theory (DMFT) shows the presence of Hunds coupling dominated description of electronic structure properties of SrCoO$_3$. This is marked by strong orbital-selective intermediate range of quasiparticle mass renormalization and significant charge fluctuations. Notably, the agreement of low temperature magnetization as well as the $T_c$ value (considering the local nature of correlation effects in single-site DMFT) with the experimental results further confirms the Hunds coupling driven magnetism. The system exhibits a largely suppressed Fermi-liquid coherence scale of temperature lower than 100 K. We argue that the presence of large incoherency in both the $e_g$ and $t_{2g}$ orbitals of Co $3d$ manifold down to lowest calculated temperature rules out the possibility of conventional double-exchange like mechanism proposed for this system in previous studies. Our study shows that the local moment picture in this system stems from a mixed spin-state configuration comprised of dynamically fluctuating IS and HS states.

## V. ACKNOWLEDGMENTS

We acknowledge the computational support provided by the High-Performance Computing (HPC) PARAM Himalaya at the Indian Institute of Technology Mandi.

# Supplementary material for "Hund's coupling driven nature of magnetism in negative charge transfer material, $SrCoO_3$"

Jyotsana Sharma[1,3],[*] Shivani Bhardwaj[1], and Sudhir K. Pandey[3][†]

[1]*School of Physical Sciences, Indian Institute of Technology Mandi, Kamand - 175075, India*
[2]*School of Mechanical and Materials Engineering,*
*Indian Institute of Technology Mandi, Kamand - 175075, India and*
[3]*Swami Vivekanand Govt. College Ghumarwin - 174021, India*
(Dated: December 14, 2025)

## I. MOMENTUM-RESOLVED SPECTRAL FUNCTION PLOT ALONG WITH THE SPIN-ORBITAL RESOLVED DFT BAND-STRUCTURE

The Momentum resolved spectral function at temperatures 110 K and 300 K is displayed in Fig.1(b)& 1(c) respectively, along with the DFT calculated band structure shown in Fig. 1(a). The spectral function shows appreciable changes with temperature considering the quasiparticle peak positions of the low-lying $e_g$ and $t_{2g}$-up (∼1-1.5 eV below the $E_f$) states. The $t_{2g}$-dn channel spans a large bandwidth ranging from ∼-0.5 to 0.5 eV around the $E_f$ and shows less thermal broadening with increase in temperature from 110-300 K, which arises from a small effective $U$ acting due to its large bandwidth, which apparently remains same as its DFT bandwidth- reflecting overall small renomalization from correlation effects and thermal broadening. Nevertheless, the loss of quasiparticle coherence is clearly visible (sharp features of spectral function signify coherency ie. quasiparticle with large lifetime, whereas, broadened curves signify incoherency i.e. quasiparticle with shorter lifetime) with increase in temperature from 110 to 300 K for all other orbitals i.e. $e_g$ manifold and $t_{2g}$-up channel, evident from narrowing of bandwidth and increase in degree of incoherency with the inclusion of correlation effects.

---

[*] jyotsana.sharma1290@gmail.com

[†] sudhir@iitmandi.ac.in

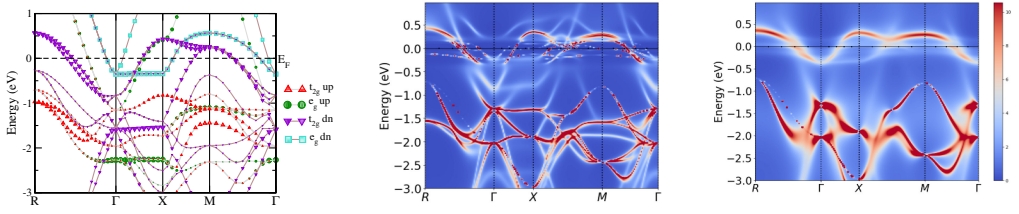

FIG. 1: (a) Orbital-resolved DFT band structure for Co-$d$. (b) Momentum-resolved spectral function at 110 K for $J$ = 1.15 eV. (c) Momentum-resolved spectral function at 300 K.